%Paper: astro-ph/9511045
%From: mannheim@main.phys.uconn.edu (Philip Mannheim)
%Date: Fri, 10 Nov 95 13:47:13 EST

\magnification=1000
%\baselineskip=0.48truecm
 \baselineskip=0.85truecm
\noindent UCONN 95-07,
November
1995
\bigskip
\centerline{\bf COSMOLOGY AND GALACTIC ROTATION CURVES}

\centerline{Philip D. Mannheim}
\centerline{Department of Physics, University of Connecticut,
Storrs, CT 06269 (permanent address), and}
\centerline{Department of Physics and Astronomy, University of
Massachusetts, Amherst, MA 01003}
\centerline{mannheim@uconnvm.uconn.edu}

\bigskip
We explore the possibility that the entire departure of galactic rotational
velocities from their luminous Newtonian expectation be cosmological in
origin,  and show that within the framework of conformal gravity (but not
Einstein  gravity apparently) every static observer sees the overall Hubble
flow as a local universal linear potential which is able to account for
available data without any need for dark matter. We find that the Universe is
necessarily an open one with 3-space scalar curvature given by
$k=-3.5\times  10^{-60}$cm$^{-2}$.
\bigskip

At the present time the search for galactic dark matter stands at an
extremely critical juncture, with neither the recent gravitational
microlensing observations of the
OGLE, MACHO and EROS collaborations
or the optical searches of the recently refurbished Hubble Space Telescope
having been able to confirm the existence of the copious
amounts of dark or faint matter that had been widely surmised to
reside in the spherical haloes of galaxies such as the Milky Way. At the
very minimum one can say that these searches have certainly not yet achieved
their intended goal of  confirming the standard
Newton-Einstein dark matter picture, while at the maximum one can  say that
they
have even thrown the entire picture into question.

Now while the standard theory is the clear preference of the bulk of the
astrophysical community, nonetheless, a small set of authors have ventured
(long before the microlensing searches in fact) to suggest that the
dark matter problem lies not in our ignorance of the matter content of
galaxies but rather in our reliance on the use of Newton's Law of Gravity on
distance scales much larger than the solar system ones on which it was first
established; with Milgrom, Bekenstein and Sanders (who all have explored the
MOND alternative) and Mannheim and Kazanas (with their conformal gravity)
having been perhaps the most persistent critics of the standard paradigm.
What mainly distinguishes the conformal gravity program (viz. gravity
based on the conformal invariant fourth order gravitational action
$I_W = -\alpha \int d^4x (-g)^{1/2} C_{\lambda\mu\nu\kappa}C^{\lambda\mu
\nu\kappa}$ where $C_{\lambda\mu\nu\kappa}$ is the conformal Weyl
tensor) from other  alternative approaches is that it sets out
to generalize not the Newtonian  potential, but rather the Schwarzschild
solution, so that from the outset  the theory is fully covariant and fully
relativistic. Indeed, exterior to a static, spherically symmetric source, the
exact vacuum metric of the theory takes the form (Mannheim and Kazanas 1989)
$ds^2=B(r)c^2dt^2- B(r)dr^2-r^2d\Omega$ where $B(r)=1-2\beta /r+\gamma r$, to
thus nicely recover both the Schwarzschild solution and its associated
Newtonian
potential, thereby enabling the theory to still meet the classic General
Relativity  tests; with the new linear potential term then only providing
corrections  to Newton on large rather than on small scales.

For galaxies, once we make the standard Newtonian assumption that the galaxies
can be treated as isolated objects (an issue we will in fact have to reconsider
below), the application of the above metric
to rotation curves is  straightforward. We simply integrate the
individual stellar potentials
$V^{*}(r)=-\beta^{*}c^2/r+ \gamma^{*}c^2r/2$ over all
the $N^{*}$ stars (and interstellar gas) in each galaxy, taking the stars to be
distributed just as the detected light and normalized
to it with a (galaxy dependent) mass to light ratio $M/L$. Now while the
ensuing
fits (see Mannheim 1993 and the more detailed unpublished fitting of Mannheim
and
Kmetko and of Carlson and  Lowenstein)  are acceptable
as far as the shape of the rotation curves is concerned
(fits in  fact almost completely identical to those of Fig. (1) which we will
present  below), the fits were only able to match the normalizations of the
curves  provided the quantity $N^{*}\gamma^{*}$ was close to a universal,
galaxy
independent value of order $10^{-30}$cm$^{-1}$. Thus rather than $\gamma^{*}$
being  found to be universal, it was the total galactic linear potential
coefficient
$\gamma_{gal}=N^{*}\gamma^{*}$ which came out universal instead. Thus unless
some reason can be found for why the effective $\gamma^{*}$ adjusts itself each
and every time to the total luminosity in each galaxy rather than the other way
round, this possible explanation of galactic rotation
curves would have to be set aside.

While the very application of $V^{*}(r)$ to the data thus shows that by itself
it does not in fact work, nonetheless we find that it fails in a very
instructive  way, namely it reveals
that the linear potential fits do capture the essence of the data, and that
they
have to be normalized in a manner which is essentially
independent of the matter content of each  individual galaxy with a magnitude
which turns out numerically to be close to  that of the inverse Hubble radius,
a
number of cosmological significance. The structure found for the fits thus
suggests that there might indeed be some universal linear  potential, but
that rather than its being due to summing over all the stars in a given galaxy,
it would instead have to be due to the effect of all of the other galaxies
in the Universe
on a given one. To  numerically motivate this possibility we calculate the
magnitude of the  centripetal acceleration
$v^2/c^2R$ at the data point farthest from the center of each of the 11 sample
galaxies listed in Table (1). (This particular set of galaxies was identified
by
Begeman, Broeils and Sanders 1991 (their paper gives complete data references)
as being a particularly reliable set of HI rotation curve data, and we use the
same galactic input parameters (distance, luminosity, stellar disk scale
length, mass of HI gas) in Table (1) as they did,    except for NGC2841 which
uses the adopted distance subsequently suggested by Sanders and Begeman 1994.)
As we can see, despite a variation of a factor of 1000 or so in luminosity, the
farthest
$v^2/c^2R$  only vary by a factor of 2 or so around a mean value of $3.1
\times  10^{-30}$cm$^{-1}$. (Because the farthest points in DDO154 may be
affected by random  gas pressures, for it we simply take the value at the point
with the highest  velocity). The data thus suggest a universal centripetal
acceleration, and thus  a universal linear potential. Indeed, independent of
our
own interest here,  this would appear to be an interesting regularity in and of
itself.

Now in the first approximation the galaxies in the Universe are distributed
smoothly in a homogeneous and isotropic Robertson-Walker (RW) geometry which
would not initially appear to have much connection to a linear potential, and
indeed in Einstein gravity there does not appear to be one as far as we can
tell.  Now the RW geometry is also a solution to the cosmology associated
with conformal gravity (Mannheim 1992). However, since the Weyl tensor vanishes
in a RW geometry (the geometry being conformal to flat), in the conformal
theory not only is RW a solution but so also is the RW metric multiplied by any
overall conformal factor (a factor which is unobservable because of the
underlying conformal invariance itself). Now in their original paper, Mannheim
and Kazanas (1989) noted the kinematic fact that under the general coordinate
transformation
$$\rho= 4r / (2(1+\gamma_0 r)^{1/2} + 2 + \gamma_0 r)~~~~~,~~~~~
\tau = \int R(t)dt \eqno(1)$$
we can effect the metric transformation
$$ds^2=(1+\gamma_0 r)c^2dt^2-dr^2/(1+\gamma_0 r)-r^2d\Omega$$
$$\rightarrow {1 \over R^2(\tau)}{(1+\rho\gamma_0/4)^2
\over (1-\rho\gamma_0/4)^2}
\left(c^2 d\tau^2 - {R^2(\tau) \over
(1-\rho^2\gamma_0^2/16)^2}
(d\rho^2 + \rho^2 d\Omega) \right) \eqno(2)$$
to yield a metric which is conformal to a RW metric with scale factor
$R(\tau)$    and (explicitly negative) 3-space scalar curvature
$k=-(\gamma_0/2)^2$. (In passing we note that in the cosmology discussed in
Mannheim 1992 an open  Universe with very negative $k$ was in fact realized,
with such a Universe not suffering from the flatness problem found in the
standard cosmology). Now, and this is the key point, in a geometry which is
both
homogeneous and  isotropic about all points, all observers can use the same
position independent conformal time
$\tau$, and any observer can serve as the origin for the coordinate
$\rho$; thus in his own local rest frame each observer is able to make the
general coordinate transformation of Eq. (1) involving his own particular
$\rho$.
Moreover, since the observer is also free in the conformal theory to make
arbitrary conformal transformations as well, that observer will then be able to
see the entire Hubble flow appear in his own local static coordinate system as
a
universal linear potential with a universal acceleration $\gamma_0c^2/2$ coming
from the spatial curvature of the Universe. Now in that specific static
coordinate system any other Hubble flow observer would see something entirely
different and not recognize anything that would look like a simple universal
linear potential at all. Only in his own explicit rest frame would any other
observer be able to recognize such a universal linear potential. However, while
the transformations of Eqs. (1) and (2) would not be useful for describing the
Hubble flow motions of the individual galaxies themselves, they appear to be
ideally suited for describing the internal orbital motions of the stars and gas
within each galaxy, since each internal motion can be discussed independently
in
each galaxy's own rest frame. Thus it would appear that in conformal gravity
each observer sees the general Hubble flow metric as a local universal linear
potential with a strength fixed by the scalar curvature of the Universe (a time
independent quantity unlike the time dependent Hubble parameter itself),
with the matter in each galaxy now acting as test particles which are being
swept
through the Hubble flow. (In passing we note the explicit role played here by
curved space. In strictly Newtonian physics the only effect of any background
would be to put tidal forces on  individual galaxies, forces that would not
account for the rotational motions of stars and gas but only to a departure
therefrom. What we find here is that the Hubble flow accounts for the explicit
motions of test particles around the center of the galaxies rather than to a
tidal perturbation to that motion. Given also the  fact that linear potentials
are not asymptotically flat, we thus see that in curved space Newtonian
reasoning can be completely misleading.)

In order to now apply this cosmological effect to explicit galactic motions, we
must look
not just at the background RW metric but rather at the embedding of each
galaxy treated as a local inhomogeneity in that background, inhomogeneities
which in the absence of the background are already putting out potentials such
as those generated by
$V^{*}$ above. Now we have not been able to find an exact solution to this
embedding problem, so for weak gravity we shall simply add the universal
cosmological  potential of Eq. (2) onto that generated by the $V^{*}$
potential.
However, if the cosmological  background is to be responsible for the
regularity
found for the galactic
$v^2/c^2R$, we should then expect the cosmological $\gamma_0$ to be of order
$10^{-30}$cm$^{-1}$. Consequently,
since $10^{12}$ or so galaxies make up the entire visible Universe, each
individual galactic
$\gamma_{gal}=\gamma^{*}N^{*}$ would then be of order
$10^{-42}$cm$^{-1}$ or so and thus completely irrelevant locally. Hence the
only
local galactic term which is relevant is the standard Newtonian one with
resulting acceleration $g_N$, with the entire local motion then being described
by
$$v^2/R=g=g_N+c^2\gamma_0/2 \eqno(3)$$
in first approximation. Now the Universe may not be exactly RW since it
appears to possess inhomogeneities on the largest scales (suggesting that
the scalar curvature may have some variation on large scales). Also in
making the transformation of Eq. (1) we ignore any local inhomogeneities
(and in particular any non-spherically symmetric ones) as well, so while we
might initially expect $\gamma_0$ to be completely universal, we note that for
fitting purposes we may anticipate some small variation in the fits.

We thus now apply Eq. (3) to our 11 galaxies, and initially fit each galaxy
with its own $\gamma_0$ and its own mass to light ratio $M/L$ to find the fits
of
Fig. (1). (The full line gives the overall fit to each galaxy, the dashed line
the
pure Newtonian contribution, and the dash dot line the pure linear
contribution).
As we can see from the fits, the linear potential model fits the data quite
well with
the pure
linear contribution completely mirroring the dark matter halo contribution
familiar
from the standard model, save only that the linear potential must eventually
cause the rotation curves to rise even while it makes then flat in the observed
region. (This key feature could eventually enable one to  distinguish between
the linear potential theory and theories such as isothermal haloes or MOND
(Milgrom 1983) which require the flatness to be an asymptotic rather than
merely an intermediate property of the rotation curves). The derived values for
$\gamma_0$ and $M/L$ are listed in Table (1), and as we see, the derived values
for the $\gamma_0$ are indeed very close,
to within a factor of 2 about a mean value. The data are thus not rejecting Eq.
(3) out of
hand. To explore the variation found for $\gamma_0$,
we also made a fit to the complete set of the 11 galaxies using just one
selfsame overall
$\gamma_0= 3.75 \times 10^{-30}$cm$^{-1}$ for the whole sample to yield the
dotted curves in Fig. (1). While not giving spot on fitting, we see that this
fitting gets to within 10\% or so of each data point, viz. much less than the
factor of 2. Finally, if we allow the adopted distances to each galaxy to vary
by up to 25\% or so we could then bring the dotted curves down to the full
curve
fits. (As such our fits are on a par with those of MOND which possesses its own
universal  acceleration $a_0$, with Begeman, Broeils and Sanders 1991 finding a
precisely similar pattern for MOND fits to the same galaxies, viz. universal
$a_0$ to within a factor of 2 galaxy by galaxy, or strictly universal $a_0$
with
instead an up to 20\% or so variation in the adopted distances.) Our fitting
should thus be regarded as acceptable and  competitive with both MOND and the
standard dark matter models. Of course, beyond the phenomenological
issue, unlike either MOND or dark matter, our  Eq. (3) is a fully motivated
output to a fully covariant theory rather than  being merely a
phenomenologically motivated input, and for that reason alone Eq. (3) is
already
to be preferred over the other contenders. Moreover, if our theory  is in fact
correct, then it provides us with an actual measurement of the scalar
curvature
of the Universe, something which despite years of intensive work has  yet to be
achieved in the standard theory.

It is of some interest to identify why it is that the linear potential
theory gives flat rotation curves at all in the observed region rather than
ones
that rise right away. The answer to this lies in a regularity first noted by
Freeman, namely that the most prominent spiral galaxies all seem to have a
common central surface brightness, $\Sigma_0^F$. (In passing we note that while
there also exist low surface brightness galaxies with  $\Sigma_0<\Sigma_0^F$,
there do not appear to be  any galaxies with $\Sigma_0>\Sigma_0^F$, thus
making
$\Sigma_0^F$ a so far unexplained upper bound on galaxies). Additionally, the
Freeman limit galaxies (galaxies which include all  the bright galaxies in our
11 galaxy sample) all seem to obey the universal   Tully-Fisher law, a
phenomenologically established universal relation between  the luminosity and
the fourth power of the velocity dispersion in the observed flat rotation curve
region. Bright galaxies thus possess a great deal of  universality. For an
exponential disk spiral ($\Sigma(R)=\Sigma_0$exp$(-R/R_0))$ we thus now note
that
since the  pure Newtonian contribution causes the rotation curve to
peak at  around $2R_0$ with a normalization which depends on
$\Sigma_0$, we can then  universally match $\gamma_0$ to
$\Sigma_0^F$ for the Freeman limit galaxies so that the value of the velocity
at
around $10R_0$ or so (a region where the linear term
dominates) will be equal  to its value at the $2R_0$ Newtonian peak in
the Freeman limit galaxies. Further, at around
$5.5R_0$ the Newtonian contribution has dropped to about half its peak value,
while the linear contribution is about half of its value at $10R_0$, to thus
give the total velocity at $5.5R_0$ a magnitude equal to its values at both
$2R_0$ and
$10R_0$, and thus a flat rotation curve from $2R_0$ all the way out to about
$10R_0$ after which the ultimate rise required of the linear potential must
begin
to set in. Unlike dark matter fits where the halo parameters have to be varied
galaxy by galaxy, in conformal gravity flatness is thus universally achieved
with no need for any adjustment of parameters. Further, since we have tuned
$\gamma_0$ to
$\Sigma_0^F$, at around $10R_0$ or so, the velocity there obeys $v^4 \sim
R_0^2(\gamma_0)^2
\sim R_0^2(\Sigma_0^F)^2 \sim \Sigma_0^F L$, which we recognize as the
Tully-Fisher relation. The universal matching of $\Sigma_0^F$ and $\gamma_0$
thus leads to both flatness and Tully-Fisher. (Since we have now
matched $\gamma_0$ to $\Sigma_0^F$, for the gas rich, low
surface brightness galaxies where $\Sigma_0<\Sigma_0^F$, it follows  that their
rotation curves should simply start rising right away, a trend which is in fact
apparent in the data. In fact this trend is the analog of the trend found
in dark matter fits where the lower luminosity galaxies are found to be
proportionately darker.) It is important to note that we have not in fact
provided an ab initio explanation for the Tully-Fisher relation since we have
not yet explained why there is in fact a Freeman limit in the first place.
However, since we have now correlated
$\Sigma_0^F$ with the cosmologically based $\gamma_0$, this suggests that the
Freeman limit may arise as upper bound on the galaxies which are generatable
as fluctuations out of the cosmological background, a background which is
indeed
controlled by the Hubble scale. The establishing of such a cosmological
origin for $\Sigma_0^F$  would then provide a complete a priori derivation
of the Tully-Fisher relation and of the systematics of rotation curves which
we have presented here.

Our uncovering of an apparent universal acceleration in conformal gravity
immediately recalls the presence of a similar feature in Milgrom's MOND,
despite
the fact that our motivation is entirely different. Specifically, Milgrom had
suggested that if a universal acceleration $a_0$ did exist, then Newton's
Second
Law could possibly be modified into a relation with a form such as
$\mu(g/a_0)g=g_N$. The candidate functional form $\mu(x)=x/(1+x^2)^{1/2}$ then
yields
$$g=g_N \{ 1/2+ (g_N^2+4a_0^2)^{1/2} / 2g_N \}^{1/2}
\eqno (4)$$
an expression which is found to perform extremely well phenomenological
despite the absence of any deeper underlying theory. Unlike Eq. (3), Eq. (4)
will lead to asymptotically flat rotation curves, and is thus quite distinct
from Eq. (3), though it is of interest to note that Eq. (3) would in fact
follow from the general MOND approach if the function $\mu(x)$ were instead to
take the form $\mu(x)= 1-1/x$. Conformal gravity thus not only provides a
rationale for why there is in fact a universal acceleration in the first
place (something simply assumed in MOND) but also yields an explicit form
for the function $\mu(x)$, albeit not the one previously considered in MOND
studies. Also of course, in conformal gravity the universal acceleration is
obtained from an equally universal (linear) potential, something which is not
the
case in MOND. Now we have noted that both Eqs. (3) and (4) both seem to fit the
available data equally well, and it would be quite remarkable if two different
formulas both worked. However, it turns out that there is a reason why they
both
work, namely the existence of the Freeman limit to which we referred above, a
limit which forces  Freeman limit galaxies with a common mass to light ratio
to
automatically  obey the  common  mass-radius relation $M=2\pi(M/L)\Sigma_0^F
R_0^2$. Now since asymptotically the MOND formula of Eq. (4) yields
$g=(MGa_0)^{1/2}/R$, the value of this quantity would then numerically
actually agree with that of Eq. (3) (viz. $g=c^2\gamma_0/2$) at around
$R=10R_0$
(i.e. at the end of the flat region in the fits of Fig. (1)) simply
because of the validity of this mass-radius relation. Since the MOND formula is
also flat all the way down to $2R_0$ for Freeman limit galaxies due to the
way the MOND function $\mu(x)=x/(1+x^2)^{1/2}$ interpolates back from the
$10R_0$ region, we see that Eqs. (3) and (4) must agree for Freeman limit
galaxies over the entire $2R_0$ to $10R_0$ region, even as they radically
disagree at larger distances. (Though we did not detect any differences in our
particular fits, detailed analysis of a larger sample of sub-Freeman limit
galaxies might eventually provide some discrimination between Eqs. (3) and
(4).)

As regards these mass-radius and Tully-Fisher relations, we
note that they apply not only to bright galaxies but also (in an
appropriate form) to globular clusters and to clusters of galaxies as well, a
feature which had been noted by Kazanas and Mannheim (1991) and examined in
detail by Schaeffer et al (1993). In particular Fig. (4) of Schaeffer et al
shows that the quantity $v^2R/L$ is essentially universal over an enormous
luminosity range from $10^4$ to $10^{12}$ solar luminosities. Thus $v^2/R \sim
L/R^2$ over the same range, i.e. the entire range is driven by effectively
a single mean surface brightness $\Sigma_0^F$ and a single universal
acceleration, to thus enable conformal gravity to immediately also give
acceptable values for the velocity dispersions in
clusters of galaxies again without needing any dark matter.
Finally, we recall that Kazanas and Mannheim (1991) also noted that the
mass-radius relation even appears to apply to the entire Universe itself
(essentially because the Schwarzschild radius of the entire Universe is of
order the Hubble radius), and, curiously even to a single elementary particle
as well. Thus we believe it is possible to make a case for the existence of a
universal linear potential associated with the cosmological Hubble flow, an
intriguing possibility which appears to eliminate the need for dark matter.

The author wishes to thank D. Kazanas, A. Broeils, S. McGaugh, S. Schneider, M.
Weinberg,  and J. Young for helpful discussions, and to thank J. Dubach for the
kind  hospitality of the Department of Physics and Astronomy of the University
of  Massachussetts at Amherst where part of this work was performed under the
New  England Land Grant Colleges Faculty Exchange Program. This work has been
supported in part by the Department of Energy under grant
No. DE-FG02-92ER40716.00.
\medskip
\noindent
{\bf References}
\smallskip
\noindent Begeman, K. G., Broeils, A. H., and Sanders, R. H. 1991,
MNRAS, 249, 523.
\smallskip
\noindent Kazanas, D., and Mannheim, P. D. 1991, Dark matter or new physics?,
in Proceedings of the ``After the First Three Minutes" Workshop, University
of Maryland, October 1990. A. I. P. Conf. Proc. No. 222, edited by
S. S. Holt, C. L. Bennett, and V. Trimble, A. I. P. (N. Y.).
\smallskip
\noindent Mannheim, P. D. 1992, ApJ, 391, 429.
\smallskip
\noindent Mannheim, P. D. 1993, ApJ, 419, 150.
\smallskip
\noindent Mannheim, P. D., and Kazanas, D. 1989, ApJ, 342, 635.
\smallskip
\noindent Milgrom, M. 1983, ApJ, 270, 365.
\smallskip
\noindent Sanders, R. H., and Begeman, K. G. 1994, MNRAS, 266, 360.
\smallskip
\noindent Schaeffer, R., Maurogordato, S., Cappi, A, and Bernardeau, F. 1993,
MNRAS, 263, L21.
\vfill\eject
\noindent
{\bf Figure Caption}
\medskip
\noindent
Figure (1). The calculated rotational velocity curves associated with the
conformal gravity potential of Eq. (3) for each of the 11 galaxies in the
sample. In each graph the bars
show the data points with their quoted
errors, the full curve shows the overall theoretical velocity prediction
(in km sec$^{-1}$) as a
function of distance from the center of each galaxy (in units of $R/R_0$
where each time $R_0$ is each galaxy's own scale length) obtained by
allowing each galaxy's acceleration parameter to vary independently, while the
dashed and dash-dotted curves show the velocities that the  Newtonian and
linear potentials would then separately produce. The dotted curve shows the
velocities that would be produced by a completely galaxy independent mean
universal acceleration.
\medskip
\noindent
{\bf Table (1)}
\smallskip
\noindent
\settabs \+Galaxy~~~~~~~~~&Distance~~& Luminosity~~&
$R_0$~~~~&$M_{HI}~~~~~~$&
M/L~~~~~~~~~~~~&$\gamma_{star}$~~~~~~~~~~~& $\gamma_{galaxy}$~~~~~~~~&
  \cr % sample line

\+ $Galaxy$& $Distance$& $Luminosity$&$R_0$&$M_{HI}$&
$v^2/c^2R$&$(M/L)_{disk}$&
$\gamma_0/2$
\cr
\medskip
\+ & $(Mpc)$& $(10^9L_{B \odot})$&$(kpc)$&$(10^9M_{\odot})$&
($10^{-30}$cm$^{-1}$)&
$(M_{\odot}/L_{B \odot})$&
 ($10^{-30}$cm$^{-1}$)
\cr
\bigskip

\+ $DDO~\phantom{1}
        154$ &        $\phantom{0}4.00$&   $\phantom{0}0.05$&   $0.50$&
$\phantom{0}0.27$&
$1.44 $&$0.60$&
       $1.26$ \cr
\smallskip
\+ $DDO~\phantom{1}
        170$ &        $12.01$&   $\phantom{0}0.16$&   $1.28$&
$\phantom{0}0.45$&
$1.63 $&$4.92$&
       $1.44 $  \cr
\smallskip
\+ $NGC~1560$&        $\phantom{0}3.00$&   $\phantom{0}0.35 $&   $1.30$&
$\phantom{0}0.82$&
$2.70$&$1.62$&
       $2.34 $  \cr
\smallskip
\+ $NGC~3109$&        $\phantom{0}1.70$&   $\phantom{0}0.81 $&   $1.55$&
$\phantom{0}0.49$&
$1.98 $&$0.03$&
       $1.63$ \cr
\smallskip
\+ $UGC~2259$&        $\phantom{0}9.80$&   $\phantom{0}1.02 $&   $1.33$&
$\phantom{0}0.43$&
$3.85 $&$3.80$&
       $2.27$ \cr
\smallskip
\+ $NGC~6503$&        $\phantom{0}5.94$&    $\phantom{0}4.80 $&  $1.73$&
$\phantom{0}1.57$&
$2.14 $&$2.96$&
       $1.97$ \cr
\smallskip
\+ $NGC~2403$&        $\phantom{0}3.25$&  $\phantom{0}7.90 $&    $2.05$&
$\phantom{0}3.10$&
$3.31 $&$1.83$&
       $2.78 $ \cr
\smallskip
\+ $NGC~3198$&        $\phantom{0}9.36$&    $\phantom{0}9.00 $&  $2.72$&
$\phantom{0}5.00$&
$2.67 $&$3.95$&
       $2.07$ \cr
\smallskip
\+ $NGC~2903$&       $\phantom{0}6.40$&    $15.30 $&   $2.02$&
$\phantom{0}2.40$&
$4.86 $&$3.53$&
       $3.80 $ \cr
\smallskip
\+ $NGC~7331$&        $14.90$&   $54.00 $&   $4.48$& $11.30$&
$5.51$&
$4.72 $&$3.73 $ \cr
\smallskip
\+ $NGC~2841$&        $18.00$&    $74.20 $&  $4.53$&   $15.90$&
$3.81 $&$2.94$&
$3.20$\cr

\end